\documentclass{aa}
\usepackage{graphicx}
\usepackage{txfonts}

\def\teff{${\rm T_{\rm eff}}$}
\def\gr{$\log {\rm g}$}
\def\vt{${\rm v_{\rm t}}$}
\def\fei{[Fe~I/H]}

\def\vk{${\rm (V-{K_s})_0}$}
\def\dteff{${\Delta \rm T_{\rm eff}}$}
\def\dgr{${\Delta \log {\rm g}}$}
\def\slope{$\sigma_{\chi}$}

\begin{document}

\title{Facing the problems on the determination of stellar temperatures and gravities: 
Galactic globular clusters
\thanks{Based on observations collected at the ESO-VLT under the programs 065.L-0507, 072.D-0507, 
073.D-0211, 078.B-0238, 081.B-0900, 083.D-0208, 085.D-0375, 089.D-0094,
093.D-0583, 095.D-0290, 188.D-3002.}}

\author{A. Mucciarelli\inst{1,2} \and P. Bonifacio\inst{3}}

\institute{
Dipartimento di Fisica e Astronomia, Universit\`a degli Studi di Bologna, Via Gobetti 93/2, I-40129 Bologna, Italy;
\and
INAF - Osservatorio di Astrofisica e Scienza dello Spazio di Bologna, Via Gobetti 93/3, I-40129 Bologna, Italy;
\and
GEPI, Observatoire de Paris, Universit\'{e} PSL, CNRS, 5, Place Jules Janssen 92195 Meudon
}

\authorrunning{Mucciarelli \& Bonifacio}
\titlerunning{Atmospheric parameters in giant stars}


\abstract{
We analysed red giant branch stars in 16 Galactic globular clusters, computing 
their atmospheric parameters both from the photometry and from excitation and ionisation balances. 
The spectroscopic parameters are lower than the photometric ones and this 
discrepancy increases decreasing the metallicity, reaching, at [Fe/H]$\sim$--2.5 dex, differences of 
$\sim$350 K in effective temperature and $\sim$1 dex in surface gravity. We demonstrate that the spectroscopic 
parameters are inconsistent with the position of the stars in the colour-magnitude diagram, providing 
too low temperatures and gravities, and predicting that the stars are up to about 2.5 magnitudes brighter 
than the observed magnitudes.

The parameter discrepancy is likely 
due to the inadequacies of the adopted physics, in particular the assumption of 1-dimensional 
geometry can be the origin of the observed slope 
between iron abundances and excitation potential that leads to low temperatures.
However, the current modelling of 3D/NLTE radiative transfer for giant stars 
seems to be not able to totally erase this slope.

We conclude that the spectroscopic parameters are wrong for metallicity lower than --1.5 dex 
and for these red giant stars photometric temperatures and gravities should be adopted. 
We provide a simple relation to correct the 
spectroscopic temperatures in order to put them onto a photometric scale.

}
 
               
\maketitle
%

\section{Introduction}
The determination of the atmospheric parameters (namely the effective temperature, \teff\ , the surface gravity, \gr\ , 
the microturbulent velocity, \vt ) is one of the most thorny aspect in the analysis of the chemical 
composition of FGK spectral type stars. 
In particular, \teff\ covers a key role because it affects any (atomic or molecular) transition, regardless of 
its ionisation stage, excitation potential, strength (at variance with \gr\ that affects mainly ionised lines 
and only marginally neutral ones, and \vt\ that affects mainly saturated lines).

\teff\ can be directly measured if the bolometric flux and the angular diameter of the 
stars are known. However, due to the sub milli-arcsecond angular size of stars, 
measures of angular diameters are restricted to a few tens of stars 
\citep[see e.g.][]{kervella04,kervella08,baines08,boyajian08,kervella17}.
Other, indirect methods to infer \teff\ have been developed, all of them undermined
by different levels of dependence by model atmospheres
(i.e. the InfraRed Flux Method, the use of Balmer line wings, the line-depth ratio).

For FGK stars \teff\ can be derived from the photometry or directly from the spectrum.
Photometric \teff\ require  dereddened broad-band colours and the adoption of suitable colour-\teff\ 
transformations \citep[see e.g.][]{alonso99,ramirez05,ghb09,casagrande10}, 
based on the InfraRed Flux Method
\citep[hereafter IRFM,][]{blackwell77, blackwell79, blackwell80}. 
This widely used approach needs of accurate/precise photometry (calibrated onto same photometric system 
where the adopted colour-\teff\ transformation is defined) and the knowledge of the colour excess, E(B-V), as well as 
information about stellar metallicity, because the colour-\teff\ relations have a mild dependence 
with [Fe/H].

One of the most popular spectroscopic methods to infer \teff\ in FGK stars 
is the so-called {\sl excitation equilibrium}, requiring 
no trend between the iron abundance A(Fe)\footnote{A(Fe)=$\log{\frac{N_{Fe}}{N_{H}}}$+12}
and the excitation potential $\chi$. Two major problems can affect \teff\ determined with this approach:\\
{\sl (i)}~low-$\chi$ lines are sensitive to \teff\ but also to 
additional effects not easy to take into account, like non local thermodynamical 
equilibrium (NLTE) and geometry/granulation effects 
\citep[see e.g.][]{bergemann12,amarsi16};\\
{\sl (ii)}~the use of spectra with a small spectral coverage, hence with a 
low number of Fe~I lines, makes uncertain the determination of the slope 
between A(Fe) and $\chi$ 
(hereafter \slope\ ), decreasing significantly the accuracy and precision in the determination of \teff\ . 
Additionally, the low-$\chi$ lines are on average the strongest ones 
\citep[see for instance Fig.~1 in][]{m13g} leading to a degeneracy between spectroscopic 
\teff\ and \vt , (the latter can be derived only spectroscopically 
by removing any trend between A(Fe) and the line strength).

Differences between the two approaches have been already highlighted in 
literature, especially in the metal-poor regime where  the
spectroscopic \teff\ turns out to be often lower than the photometric ones by some hundreds of K 
\citep[see e.g.][]{johnson02,cayrel04,cohen08,frebel13,yong13}.
Such differences can lead to lower absolute abundances (by $\sim$0.2-0.3 dex), 
can falsify the abundance ratios and 
introduce systematics that erase the precision due to the spectral quality.

In this work we analyse a representative sample of red giant branch (RGB) stars in 
16 Galactic globular clusters (GCs) 
with the aim to compare parameters derived from the spectroscopic and photometric approaches described 
above and highlight possible bias in the two methods.
GCs are powerful tools to perform this kind of comparison, because 
colour excess (fundamental to derive photometric \teff\ ),  
distance and stellar mass (necessary to calculate \gr\ ) 
can be easily obtained from the isochrone-fitting of the main sequence turnoff point 
observed in their colour-magnitude diagram (CMD). 
Also, because a GC can be efficiently described 
as a single-age, single-metallicity population, 
its RGB stars follow a well-defined \teff\--\gr\ relation, described by the theoretical isochrone 
with the corresponding cluster age and metallicity, thus providing a solid, physical reference to evaluate 
the reliability of the adopted parameters.


\section{The spectroscopic dataset}
We selected 16 Galactic GCs with the following criteria:\\
{\sl (i)}~clusters covering the entire metallicity range of the Galactic halo GC system, 
between [Fe/H]$\sim$--2.5 dex and $\sim$--0.7 dex;\\
{\sl (ii)}~GCs with available archival spectra secured with UVES-FLAMES 
mounted at the Very Large Telescope of the European Southern Observatory. 
This spectrograph provides a high spectral resolution (R=47,000) and a wide 
spectral coverage (Red Arm 580, 4800-6800 \AA\ ), thus providing a large number of Fe~I and 
Fe~II lines, necessary to robustly derive spectroscopic atmospheric parameters.
Note that a huge sample of spectra of GC stars is available with the multi-object spectrograph 
GIRAFFE-FLAMES@VLT but these spectra, because of the limited spectral coverage, 
can not be suitable to robustly derive spectroscopic parameters, in particular for the most metal-poor GCs, 
due to the low number of Fe~I lines (especially those with low $\chi$) and the lack of Fe~II lines;\\
{\sl (iii)}~GCs with available ground-based UBVI photometry from the database maintained 
by P. B. Stetson\footnote{http://www.cadc-ccda.hia-iha.nrc-cnrc.gc.ca/en/community/\\STETSON/homogeneous/}
\citep[see][]{stetson19} and calibrated in the standard \citet{landolt92} photometric system. 
For these clusters J$K_S$ near-infrared photometry 
is available from the Two Micron All Sky Survey \citep[2MASS,][]{2mass};\\
{\sl (iv)}~GCs with relatively low colour excess (E(B-V)$<$0.2 mag) in order to minimise the 
effects of differential reddening that can reduce the precision in the photometric parameters.

\begin{table}
\caption{Spectroscopic dataset of the target globular clusters (sorted 
in increasing metallicity), including the number of analysed stars, the colour excess, the 
V-band distance modulus and the corresponding ESO program.}             
\label{parcl}      
\centering                          
\begin{tabular}{c c c c c}        
\hline\hline                 
CLUSTER & ${\rm N_{stars}}$ & E(B-V) & ${\rm (m-M)_V}$ & Program \\    
\hline
   &   &  (mag)  &  (mag)  &  \\
\hline                        
  NGC~7078  &  13  &  0.090  &  15.44   &  073.D-0211 \\     
  NGC~4590  &  13  &  0.065  &  15.23   &  073.D-0211 \\   
  NGC~7099  &  19  &  0.050  &  14.78   &  073.D-0211 \\    
            &      &         &          &  085.D-0375 \\    
  NGC~6397  &  12  &  0.195  &  12.63   &  073.D-0211   \\ 
  NGC~5694  &   6  &  0.110  &  18.25   &  089.D-0094 \\   
  NGC~5824  &   6  &  0.140  &  17.95   &  095.D-0290 \\   
  NGC~5634  &   7  &  0.060  &  17.20   &  093.B-0583  \\   
  NGC~6809  &  13  &  0.120  &  14.00   &  073.D-0211 \\   
  NGC~6093  &   9  &  0.200  &  15.76   &  083.D-0208   \\ 
  NGC~1904  &  10  &  0.030  &  15.60   &  072.D-0507 \\   
  NGC~6752  &  12  &  0.070  &  13.27   &  073.D-0211 \\   
  NGC~288   &  10  &  0.015  &  14.83   &  073.D-0211 \\   
  NGC~5904  &  14  &  0.030  &  14.43   &  073.D-0211 \\   
  NGC~1851  &  23  &  0.015  &  15.37   &  188.B-3002 \\   
  NGC~2808  &  12  &  0.170  &  15.55   &  072.D-0507 \\   
  NGC~104   &  10  &  0.045  &  13.44   &  073.D-0211 \\   
\hline                                   
\end{tabular}
\end{table}

\begin{table*}
\caption{Average iron abundances for the target clusters derived from 
photometric parameters (from Fe~I and Fe~II lines) and from 
spectroscopic parameters (from Fe~I lines).}            
\label{ironcl}      
\centering                         
\begin{tabular}{ccccccc}        
\hline\hline                 
CLUSTER & [Fe~I/H] & $\sigma$ & [Fe~II/H] & $\sigma$ & [Fe~I/H] & $\sigma$ \\
\hline
  & (PHOTOM) &  & (PHOTOM) &  & (SPEC)  \\
\hline
   &  (dex) &  (dex)  &  (dex) & (dex)  & (dex) & (dex) \\
\hline   
 NGC~7078    &   --2.42   &    0.07   &    --2.40   &    0.04   &   --2.71    &    0.09      \\
 NGC~4590    &   --2.28   &    0.05   &    --2.31   &    0.05   &   --2.60    &    0.07      \\
 NGC~7099    &   --2.31   &    0.05   &    --2.32   &    0.05   &   --2.61    &    0.07      \\
 NGC~6397    &   --2.01   &    0.03   &    --2.07   &    0.04   &   --2.25    &    0.06      \\
 NGC~5694    &   --1.92   &    0.05   &    --2.03   &    0.09   &   --2.11    &    0.08      \\
 NGC~5824    &   --1.92   &    0.05   &    --2.00   &    0.05   &   --2.08    &    0.05      \\
 NGC~5634    &   --1.80   &    0.05   &    --1.87   &    0.03   &   --1.96    &    0.04      \\
 NGC~6809    &   --1.73   &    0.03   &    --1.81   &    0.03   &   --1.90    &    0.04      \\
 NGC~6093    &   --1.77   &    0.03   &    --1.78   &    0.02   &   --1.80    &    0.04      \\
 NGC~1904    &   --1.52   &    0.03   &    --1.56   &    0.01   &   --1.62    &    0.03      \\
 NGC~6752    &   --1.49   &    0.03   &    --1.66   &    0.03   &   --1.62    &    0.03      \\
  NGC~288    &   --1.23   &    0.04   &    --1.39   &    0.06   &   --1.29    &    0.03      \\
 NGC~5904    &   --1.22   &    0.03   &    --1.31   &    0.05   &   --1.24    &    0.03      \\
 NGC~1851    &   --1.12   &    0.03   &    --1.16   &    0.04   &   --1.13    &    0.04      \\
 NGC~2808    &   --1.06   &    0.07   &    --1.18   &    0.07   &   --1.09    &    0.06      \\
  NGC~104    &   --0.75   &    0.03   &    --0.76   &    0.04   &   --0.76    &    0.03      \\

 \hline                                  
\end{tabular}
\end{table*}

\begin{table*}
\caption{Average differences between spectroscopic and photometric parameters for each target cluster. For each value the corresponding dispersion of the mean is listed.}            
\label{ironcl2}      
\centering                         
\begin{tabular}{c ccccc ccccc}        
\hline\hline                 
CLUSTER  & 
$\Delta{\rm T_{eff}}$ & $\sigma$ &
$\Delta{\rm log~g}$ & $\sigma$ &
$\Delta{\rm v_{t}}$ & $\sigma$ &
$\Delta{\rm [Fe/H]}$ & $\sigma$ &
\\   
\hline
    & (K) & (K) &   (dex)     &     (dex)    &  (km/s) & (km/s) & 
   (dex) & (dex) \\
\hline   
 NGC~7078     &    --330  &         56  &   --1.01   &   0.20   &  --0.16   &   0.11   &  --0.29	 &  0.06   \\
 NGC~4590     &    --365  & 	    40  &   --1.08   &   0.13   &  --0.29   &   0.08   &  --0.32	 &  0.04   \\
 NGC~7099     &    --352  & 	    72  &   --1.06   &   0.14   &  --0.24   &   0.14   &  --0.30	 &  0.07   \\
 NGC~6397     &    --247  & 	    42  &   --0.58   &   0.15   &  --0.18   &   0.17   &  --0.24	 &  0.05   \\
 NGC~5694     &    --193  & 	    33  &   --0.42   &   0.16   &  --0.20   &   0.06   &  --0.18	 &  0.04   \\
 NGC~5824     &    --153  & 	    23  &   --0.47   &   0.06   &  --0.15   &   0.10   &  --0.17	 &  0.03   \\
 NGC~5634     &    --174  & 	    28  &   --0.49   &   0.13   &  --0.11   &   0.04   &  --0.16	 &  0.03   \\
 NGC~6809     &    --160  & 	    26  &   --0.42   &   0.12   &  --0.05   &   0.07   &  --0.17	 &  0.03   \\
 NGC~6093     &     --38  & 	    35  &   --0.12   &   0.10   &  --0.01   &   0.03   &  --0.03	 &  0.03   \\
 NGC~1904     &    --111  & 	    31  &   --0.32   &   0.12   &  --0.05   &   0.07   &  --0.10	 &  0.03   \\
 NGC~6752     &    --153  & 	    40  &   --0.13   &   0.09   &  --0.08   &   0.06   &  --0.13	 &  0.04   \\
  NGC~288     &     --85  & 	    23  &    +0.02   &   0.10   &  --0.02   &   0.06   &  --0.05	 &  0.04   \\
 NGC~5904     &     --19  & 	    32  &    +0.09   &   0.08   &   +0.02   &   0.07   &  --0.02	 &  0.04   \\
 NGC~1851     &     --25  & 	    34  &   --0.01   &   0.12   &  --0.01   &   0.06   &  --0.01	 &  0.02   \\
 NGC~2808     &     --47  & 	    40  &    +0.09   &   0.12   &   +0.00   &   0.06   &  --0.02	 &  0.04   \\
  NGC~104     &     --57  & 	    23  &    +0.01   &   0.11   &   +0.03   &   0.05   &  --0.01	 &  0.04   \\

 \hline                                  
\end{tabular}
\end{table*}

\section{Determination of the atmospheric parameters}

\subsection{Photometric parameters}
\label{photpars}
Photometric \teff\ have been derived through the colour-\teff\ transformations by \citet{ghb09} 
that provide relations for giant stars for the broad-band colours
${\rm (B-V)_0}$, \vk\ and ${\rm (J-{K_s})_0}$. 
These dereddened colours have been obtained with the BV magnitudes from \citet{stetson19} and 
the near-infrared J${\rm K_s}$ magnitudes from the 2MASS database
\citep{2mass}, adopting the extinction coefficients by \citet{mccall04}.
Because the 2MASS magnitudes  have  uncertainties larger than the optical magnitudes, 
especially for the farther clusters, 
we adopted as J and $K_s$ magnitudes those obtained by projecting the position of each individual 
star on the mean ridge line of the RGB in the (${\rm K_s}$, J-${\rm K_s}$) CMD.

Colour excess E(B-V) and V-band distance modulus ${\rm (m-M)_V}$ for each cluster have been 
estimated through a best-fit of the (V, V-I) CMD with 
theoretical isochrones from the DARTMOUTH Stellar Evolution Database \citep{dotter08}. 
The derived values of E(B-V) and ${\rm (m-M)_V}$ for each target cluster are listed in Table~\ref{parcl}.
These values have been compared 
with those listed by \citet{harris}:  
we found average differences between 
our values and those by \citet{harris} of +0.008 mag ($\sigma$=~0.02 mag) and 
+0.07 mag ($\sigma$=~0.10 mag) for 
E(B-V) and ${\rm (m-M)_V}$, respectively. We stress that our values have been determined in an homogeneous way, while \citet{harris} presents a compilation of values derived from different sources and methods.

Surface gravities have been estimated adopting the photometric \teff\ , 
a stellar mass obtained from the corresponding best-fit theoretical isochrone and 
the bolometric corrections calculated with the relations provided by \citet{alonso99}.
Microturbulent velocities have been estimated by erasing any trend between 
iron abundances and reduced EWs (defined as $\log{\frac{EW}{\lambda}}$) . 
Only for the cluster NGC~2808, the photometric catalog has been corrected for 
differential reddening.

\subsection{Spectroscopic parameters}
In this approach, all the stellar parameters have been estimated 
from the spectra, requiring to fulfil three constraints 
\citep[see][for a detailed description of the procedure adopted here]{m13g}:\\
{\sl (i)}~\teff\ are obtained from the so-called 
{\sl excitation equilibrium}, requiring no trend between 
iron abundances and $\chi$ (\slope\ $\sim$0);\\
{\sl (ii)}~\gr\ are obtained from the so-called 
{\sl ionisation equilibrium}, requiring that neutral and single 
ionised Fe lines provide the same average abundance, within the corresponding uncertainties 
([Fe~I/Fe~II]$\sim$0);\\
{\sl (iii)}~\vt\ are obtained with the same 
approach described above for the photometric parameters.

Thanks to their high spectral quality (signal-to-noise ratio$>$100),
the spectra analysed in this work allow to measure 
$\sim$100-200 Fe~I lines in each star (depending on the metallicity), well distributed both in reduced EW and 
$\chi$, and 10-20 Fe~II lines, providing a robust statistical ground to use this approach.

\section{Chemical analysis}
The chemical abundances and the spectroscopic atmospheric parameters 
have been obtained with the package {\tt GALA} \citep{m13g} that calculates 
the abundances by matching observed and theoretical EWs of unblended lines.

Neutral and single ionised iron lines have been selected by comparing any observed 
spectrum with a synthetic spectrum calculated with the corresponding photometric 
parameters and assuming the cluster iron abundances listed by \citet{harris} as guess values. 
Synthetic spectra have been calculated with the code {\tt SYNTHE} 
\citep{sbordone04,kurucz05}, including all the atomic and molecular 
transitions available in the Kurucz/Castelli 
database\footnote{http://wwwuser.oats.inaf.it/castelli/linelists.html}.

Plane-parallel, 1-dimensional model atmospheres have been calculated 
for each star with the {\tt ATLAS9} code \citep{kurucz05} adopting 
local thermodynamical equilibrium (LTE) for all species 
and without the use of the {\sl approximate overshooting}.
All the model atmospheres 
have been computed by interpolating at the cluster metallicity the opacity distribution 
functions by \citet{castelli03}, adopting for all the clusters 
an $\alpha$-enhanced chemical mixture but for NGC~5694 for which 
a solar-scaled chemical mixture is adopted \citep{m5694}.

Laboratory oscillator strengths for Fe~I lines are from \citet{martin88} and \citet{fuhr06}. 
At variance with Fe~I lines, few of Fe~II lines have laboratory oscillator strengths and, 
even if they are accurate, they are often imprecise, with large uncertainties 
\citep[see e.g. the critical discussions about the {\sl} gf-values of Fe~II lines in \citet{lambert96} and][]{mel09}. 
For single-ionised Fe lines we adopted the 
oscillator strengths by \citet{mel09} that included theoretical {\sl gf}-values, 
with high precision for the components of the same multiplet, 
calibrated on laboratory data or on the solar spectrum.

EWs have been measured using the {\tt DAOSPEC} code \citep{stetson08}
managed through the wrapper {\tt 4DAO} \citep{4dao}.

Strong lines, located in the flat part of the curve of growth have been 
excluded because they are sensitive to the velocity fields 
and less sensitive to the abundance. 
The threshold in reduced EW has been chosen 
depending on the cluster metallicity 
(higher the metallicity/lower the temperature larger the reduced EW 
corresponding to the starting point of the flat part of the
curve of growth). 
Additionally, weak (noisy) lines have been excluded, as well as 
lines with discrepant abundances with respect to the abundance 
distribution from the other lines.

\section{Results}

\subsection{Spectroscopic vs. photometric parameters}

Table~\ref{ironcl} lists the average abundances for each target cluster 
obtained adopting photometric (from Fe~I and Fe~II lines) 
and spectroscopic parameters (from Fe~I lines only). For each abundance ratio 
the dispersion of the mean is quoted. All the clusters, regardless of the 
adopted set of parameters, exhibit small dispersions of the mean, 
reflecting their high level of intrinsic homogeneity of the metallicity  \citep[see e.g.][]{c09iron}. 

Table~\ref{ironcl2} lists for each target cluster the average differences between the spectroscopic 
and photometric parameters with the corresponding dispersion of the mean.

\begin{figure}
\includegraphics[width=\columnwidth]{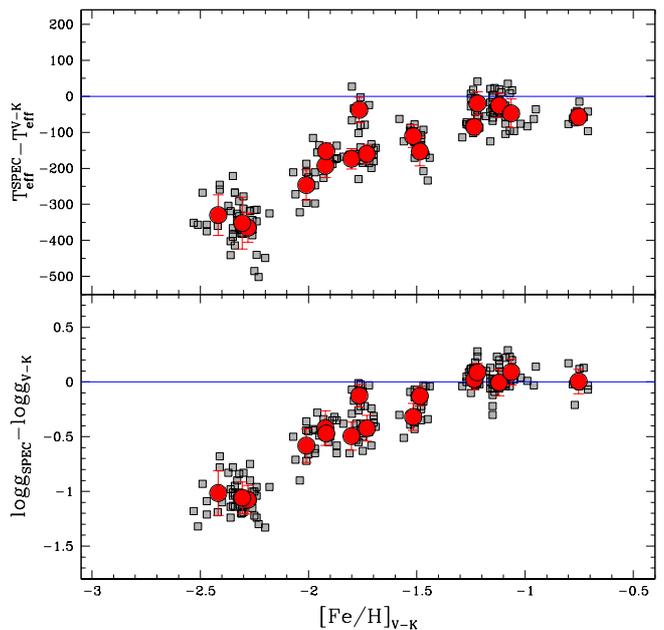}
\caption{Behaviour of the difference between 
spectroscopic and \vk\--based \teff\  (upper panel) and 
\gr\ (lower panel) as a function 
of the iron abundance [Fe/H] derived from the photometric parameters,
for individual stars (small grey squares) 
and average values for each cluster (red points);
 the errorbars indicate the 1-$\sigma$ dispersion by the mean.}
\label{vialo}
\end{figure}

Fig.~\ref{vialo} shows the run of the difference between spectroscopic and 
photometric \teff\ and \gr\ (upper and lower panel, respectively) as a function 
of [Fe/H], the latter obtained with the photometric parameters. 
In both panels grey squares mark individual cluster stars, while 
large red circles are the average value for each cluster 
(the errorbars indicate 1-$\sigma$ dispersion of \dteff\ and \dgr\ ). 
In this figure we adopted the stellar parameters and metallicities 
derived from \vk\ , one of the most used \teff\ indicators because of its wide colour baseline 
(as discussed in Section~\ref{checks}, the other colours 
provide the same behaviour shown in Fig.~\ref{vialo}).

\begin{figure}
\includegraphics[width=\columnwidth]{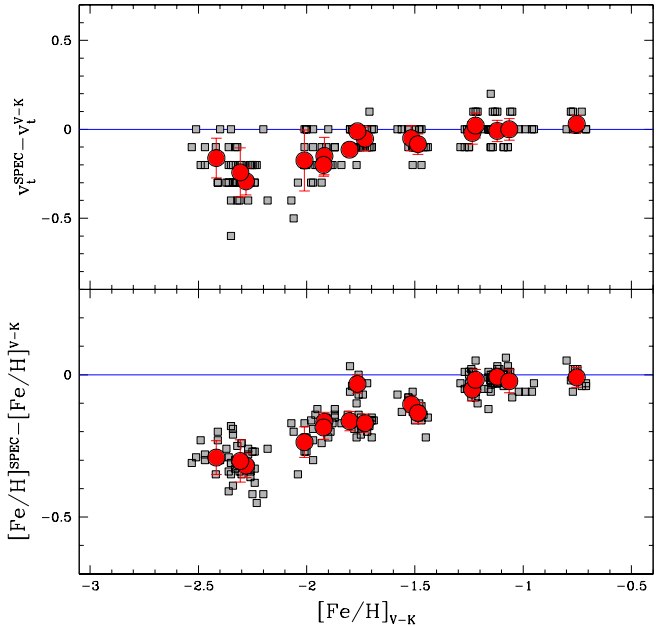}
\caption{Behaviour of the difference between 
spectroscopic and \vk\--based \vt\  (upper panel) and 
[Fe/H] (lower panel) as a function 
of the iron abundance [Fe/H] derived from the photometric 
parameters (same symbols of Fig.~\ref{vialo}).}
\label{vialo2}
\end{figure}

Fig.~\ref{vialo2} shows the run of the difference between spectroscopic and 
photometric \vt\ and [Fe/H] (upper and lower panel, respectively) as a function 
of [Fe/H] , adopting the same symbols of Fig.~\ref{vialo}.

The differences between spectroscopic and photometric parameters exhibit a clear 
run with the metallicity, with spectroscopic \teff\ , \gr\ and \vt\ that 
decrease with respect to the photometric values moving toward lower metallicities. 
In particular, for GCs with [Fe/H] $>$--1.3 dex 
(namely NGC~288, NGC~5904, NGC~1851, NGC~2808 and NGC~104) 
the two sets of parameters agree very well each other, with an average offset 
of about --50 K for \teff\ , +0.04 for \gr\ and 
+0.01 km/s for \vt\ . These differences lead to negligible changes in the derived metallicities.

For the GCs with [Fe/H] between --2.0 dex and --1.5 dex
the spectroscopic parameters are lower than the photometric ones by $\sim$100-200 K for \teff\ , 
--0.1/--0.5 for \gr\ and 
 $\sim$0.0/--0.3 dex for \vt\ . 
 The [Fe/H] derived from spectroscopic parameters 
 is about 0.15-0.2 dex lower than those obtained with the 
 photometric values.

Finally, for the most metal-poor clusters of the sample 
(namely NGC~7078, NGC~4590 and NGC~7099) the spectroscopic \teff\ are 
lower by $\sim$350 K, the spectroscopic \gr\  
are lower by 1 dex and \vt\ differ of $\sim$0.3 km/s. 
The iron abundances derived from spectroscopic 
parameters are lower by $\sim$0.3 dex with respect to those 
obtained with photometric parameters.

The lower spectroscopic \teff\ obtained for the most metal-poor clusters arise 
from the significant \slope\ found when photometric \teff\ are adopted.

\begin{figure}
\includegraphics[width=\columnwidth]{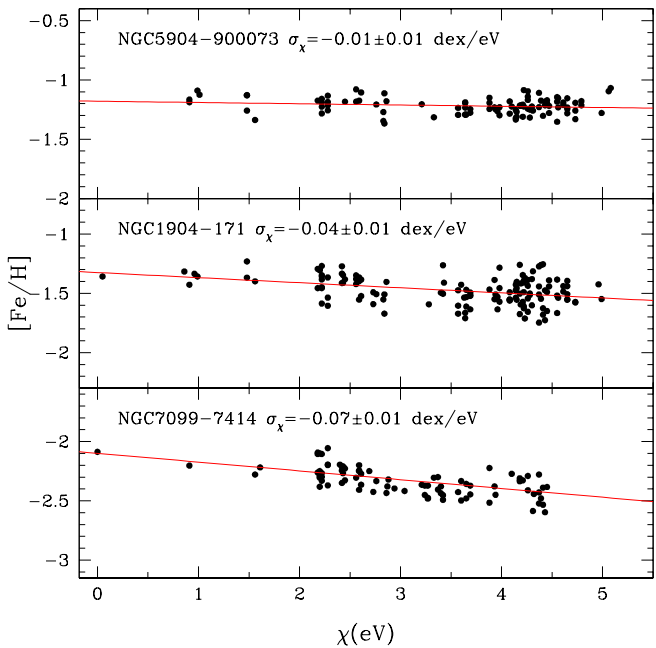}
\caption{Behaviour of [Fe/H] as a function of $\chi$ for 
individual Fe~I lines in three stars in NGC~5904 (upper panel), 
NGC~1904 (middle panel) and NGC~7099 (lower panel), adopting photometric parameters. 
Red lines are the linear best-fits.
The slopes between [Fe/H] and $\chi$ are labelled. }
\label{exa1}
\end{figure}

As an example of the measured \slope\ , 
Fig.~\ref{exa1} shows the behaviour of Fe~I abundances 
as a function of $\chi$ for three stars in NGC~5904, NGC~1904 and 
NGC~7099 when the \vk\ -based \teff\ are adopted. 
The values of \slope\ for the metal-rich clusters are compatible 
with a null slope and they become 
more negative decreasing the metallicity, reaching values of --0.07/--0.10 
dex/eV for the three most metal-poor target clusters.

\subsection{Sanity checks} 
\label{checks}
We performed some sanity checks to assess the validity of the trends discussed above.

\begin{itemize}
\item 
We repeated the analysis using photometric \teff\ 
derived from the ${\rm (B-V)_0}$ and ${\rm (J-K)_0}$-\teff\ transformations provided 
by \citet{ghb09}.
The average differences between spectroscopic \teff\ and those obtained from these two additional 
colours are shown in Fig.~\ref{other} as red circles: the behaviour with the metallicity well 
resembles that obtained with \vk\ .
{\sl The observed run of the differences of the parameters with the metallicity is independent of the adopted colours.}\\

\item We checked whether the observed trend is due to the adopted 
colour-\teff\ transformations by \citet{ghb09}. 
We re-analysed the target stars using the \citet{alonso99} relations that provide 
colour-\teff\ transformations for ${\rm (U-V)_0}$, ${\rm (B-V)_0}$, ${\rm (V-I)_0}$, \vk\ and ${\rm (J-{K_s})_0}$.
We adopted the extinction coefficients by \citet{mccall04}, the 
optical UBVI magnitudes from \citet{stetson19} and 
the near-infrared J${\rm K_s}$ magnitudes from the 2MASS survey
\citep{2mass}. The latter have been transformed into {\sl Telescopio Carlos Sanchez} photometric 
system, adopted by \citet{alonso99}, using the relations by \citet{carpenter01}. 
The results are shown in Fig.~\ref{other} as blue triangles.
The \citet{alonso99} scale is cooler than that by \citet{ghb09} by 47 K ($\sigma$=~35 K), 
105 K ($\sigma$=~11 K) and 83 K ($\sigma$=~17 K) for ${\rm (B-V)_0}$, \vk\ and ${\rm (J-{K_s})_0}$, 
respectively.
Despite these differences between the two scales, the same behaviour with the 
metallicity is found, indicating that {\sl this run is not an artefact of the 
adopted \teff\ scale.}\\

\begin{figure}
\includegraphics[width=\columnwidth]{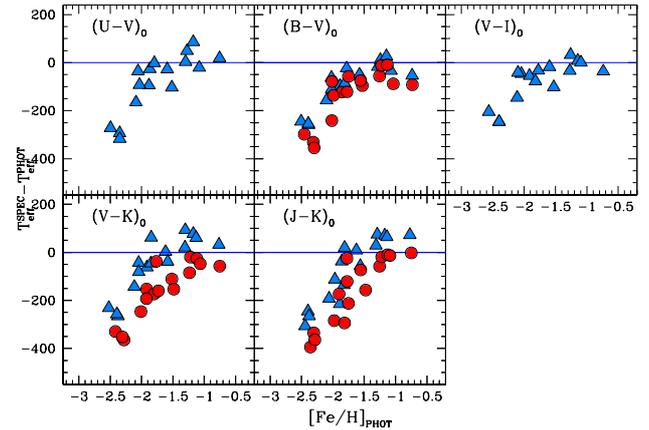}
\caption{As Fig.~\ref{vialo} adopting 
the relations by \citet[][blue triangles]{alonso99} 
for the colours ${\rm (U-V)_0}$, ${\rm (B-V)_0}$, ${\rm (V-I)_0}$, ${\rm (V-{K_s})_0}$ and ${\rm (J-{K_s})_0}$, 
and those by \citet[][red circles]{ghb09} for the colours ${\rm (B-V)_0}$, ${\rm (V-{K_s})_0}$ and ${\rm (J-{K_s})_0}$. 
Only the average values for each target cluster are shown 
and not the individual stars.}
\label{other}
\end{figure}

\item The target stars have been re-analysed by excluding 
Fe~I lines with $\chi$ $<$ 2 eV. These lines are more affected 
by inadequacies in the model atmospheres, in particular 
3D effects \citep{bergemann12,dobrovolskas13}.
A similar selection has been already adopted in other 
studies, even if with different thresholds 
\citep[see e.g.][]{cayrel04,cohen08,yong13,ruchti13}. 
Spectroscopic \teff\  derived ruling out the low-energy transitions 
continue to be significantly lower (by $\sim$200-300 K) than the photometric 
ones, for stars with \fei\ $<$--2.0 dex. 
As clearly visible in the lower panel of Fig.~\ref{exa1}, 
significant values of \slope\ ($\sim$0.07-0.10 dex/eV) are found in metal-poor stars 
also when the low-energy Fe~I lines are excluded.
The inclusion of a ten of low-$\chi$ lines with 
higher abundances decreases \vt\ by $\sim$0.3 km/s 
(because the excluded transitions are on average the 
strongest ones and hence the more sensitive to \vt\ ) but 
has a negligible impact on the average \fei\ , because of 
the large number of high-energy lines in our linelist. \\

\item 
The determination of spectroscopic 
\gr\ can be affected by the choice of the 
used {\sl gf}-values of the Fe~II lines, because the latter are less precise than 
those available for Fe~I lines and 
laboratory values are available only for a few lines.
We checked two alternative sets of Fe~II {\sl gf}-values, 
the laboratory oscillator strengths 
provided by \citet{kroll87,heise90} and \citet{hannaford92} 
and the theoretical ones by \citet{raa}.

Gravities obtained assuming the laboratory 
values are compatible, within the uncertainties, with 
those obtained adopting the values by 
\citet{mel09}, with an average difference 
(laboratory - this work) of 
--0.03 dex ($\sigma$=~0.09 dex). 
On the other hand, gravities obtained with 
theoretical {\sl gf}-values are lower than 
our ones, with an average difference 
of --0.28 dex ($\sigma$=~0.13 dex), 
increasing the discrepancy between 
the photometric and spectroscopic \gr\ .
In both cases, spectroscopic \teff\ and \vt\ are not affected by the choice of the 
{\sl gf}-values of Fe~II lines. 

These checks demonstrate that the strong difference 
between spectroscopic and photometric \gr\ shown in Fig.~\ref{vialo} cannot be 
attributed to the uncertainties in the 
adopted {\sl gf}-values of Fe~II lines.\\

\item Finally we checked whether the adoption of a different combination 
of model atmospheres/spectral synthesis code can alleviate or solve 
the observed discrepancies. We repeated the analysis 
using the code {\tt TURBOSPECTRUM} \citep{plez12} 
coupled with the MARCS model atmospheres \citep{marcs08} but this choice 
does not change the observed runs.

\end{itemize}


\section{Previous works}

This work presents for the first time a homogeneous comparison between 
the two approaches used to derive stellar parameters and performed 
on the entire metallicity range of the Galactic GCs. 
The analysis of individual metal-poor clusters is not sufficient to 
highlight the overall behaviour that we have identified because the difference 
between photometric and spectroscopic parameters can be interpreted as a systematics (and not 
as a metallicity dependent phenomenon).
However, hints of a similar behaviour have been found in 
some previous works analysing 
GCs with different metallicities.

\citet{c09} analysed 202 giant stars in 17 GCs observed with 
UVES-FLAMES@VLT, adopting photometric parameters and finding that \slope\ turn out to be 
more negative in metal-poor stars. They provided 
an average slope \slope\ =--0.013 dex/eV ($\sigma$=~0.029 dex/eV) 
suggesting that the photometric \teff\ should be decreased 
by 45 K to obtain an average, null \slope\ .
This approach interprets the average slope as the result of 
an offset between spectroscopic and photometric \teff\ , 
while the effect becomes significant only at low metallicity. 
In particular, \citet{c09} found for the GCs with [Fe/H]$<$--2.0 dex 
slopes of --0.04/--0.07 dex/eV ; these values are higher than those 
obtained in our work with the \citet{ghb09} relations but compatible 
with those that we found with the \citet{alonso99} transformations 
\citep[the same used by][]{c09}.

\citet{nidever19} in their study on the chemical composition 
of Magellanic Clouds giant stars compared the iron content
of 14 southern Galactic clusters observed with APOGEE-2S 
with the values listed by \citet{c09}. We recall that the analysis 
by \citet{c09} is based on photometric parameters while the 
atmospheric parameters for the APOGEE-2S targets have been 
derived from the {\tt ASPCAP} pipeline \citep{aspcap} by fitting the 
observed spectra with synthetic ones in specific spectral regions sensitive to any parameters
(hence, they are spectroscopic parameters even if they have been derived 
with a different approach with respect to that used here). 
The agreement is satisfactory down to [Fe/H]$\sim$--2.0 dex, 
while for the most metal-poor clusters the iron abundance 
for the spectroscopic parameters by ASPCAP is lower by 
$\sim$0.2 dex than the iron content derived by \citet{c09}.
Because of the different wavelength range and the use of different 
diagnostics, the origin of this discrepancy between optical and near-infrared 
GC metallicities is not trivial to disentangle, in particular because 
no comparison about \teff\ and \gr\ for the stars in common 
is discussed. However, a more accurate comparison between the two analyses should be performed to understand whether the discrepancy highlighted by \citet{nidever19} for the metal-poor GCs is due to the different methods to estimate the atmospheric 
parameters or other effects related to the different spectral ranges.
For a more detailed comparison between 
near-infrared spectroscopic and photometric 
\teff\ we refer the reader to \citet{mes15,jonsson18,masseron19} and \citet{mes20}.

\citet{kovalev19} analysed some open and globular clusters determining stellar 
parameters by comparing the observed spectra with both LTE and NLTE synthetic spectra. 
The differences in \teff\ and \gr\ between the LTE and NLTE analyses, both 
based on spectroscopic diagnostics and not photometric parameters, is qualitatively 
analogue to those obtained in this work; in particular, for clusters with [Fe/H]$<$--2.0 dex, 
they found that LTE spectroscopic \teff\ and \gr\ are lower than the NLTE ones by $\sim$200-300 K 
and 0.4-0.6 dex, respectively.

\section{Discussion}

\subsection{Which parameter set should be preferred for metal-poor giant stars?}

As explained in Section 1, one of the main advantages to work 
with GCs is that we can easily compare the  
atmospheric parameters (derived from photometry or spectroscopy) 
with the values predicted by appropriated theoretical isochrones. 
This provides a powerful and exemplary check to decide whether a 
given set of parameters is correct or not, because it should be 
consistent with the position of the star in the CMD.

\begin{figure*}
\centering
\includegraphics[width=17cm]{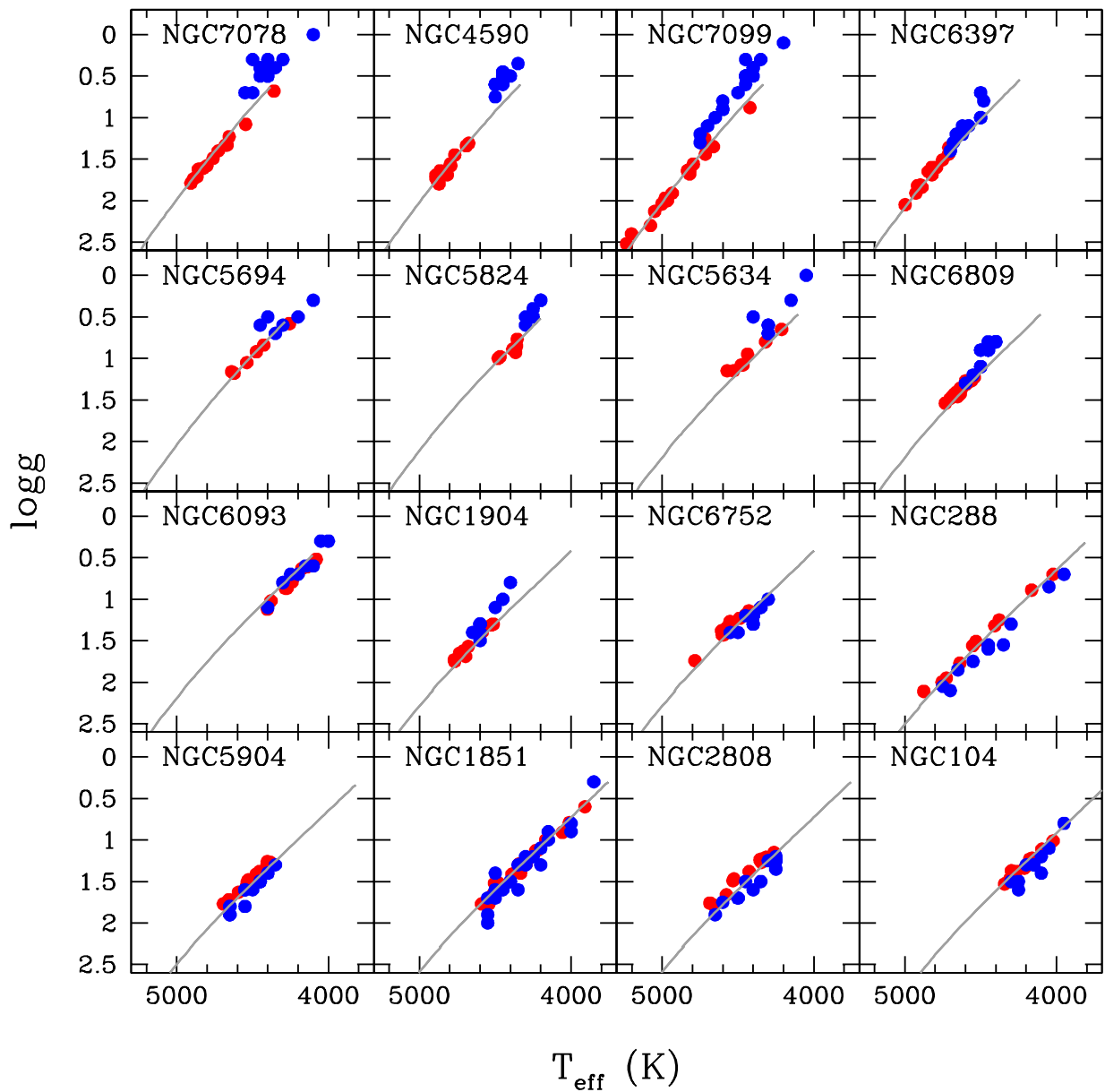}
\caption{Behaviour of \gr\  as a function of \teff\ 
for all the target clusters (sorted for increasing metallicity);
blue points are the spectroscopic parameters and red points 
the photometric ones. For each cluster the corresponding best-fit DARTMOUTH theoretical 
isochrone is shown (grey solid line).}
\label{isoc}
\end{figure*}

Fig.~\ref{isoc} shows the position of the individual GC stars in the 
\teff - \gr\ diagram (red and blue circles are the photometric and 
spectroscopic parameters, respectively), with superimposed the 
corresponding best-fit theoretical DARTMOUTH isochrone. 
The photometric parameters well match with the those predicted by 
the isochrones.
On the other hand, the position of the stars, when the spectroscopic 
parameters are adopted, shifts systematically toward lower \teff\ and \gr\ decreasing the cluster metallicity.

A simple argument to prefer photometric parameters is that 
the spectroscopic ones predict for these stars a wrong position 
in the \teff\--\gr\ diagram.
For most of the investigated GCs, the target stars are 1-3 magnitudes 
fainter than the RGB Tip, while the spectroscopic parameters locate 
them close to the Tip of the RGB. 

\begin{figure}
\includegraphics[width=\columnwidth]{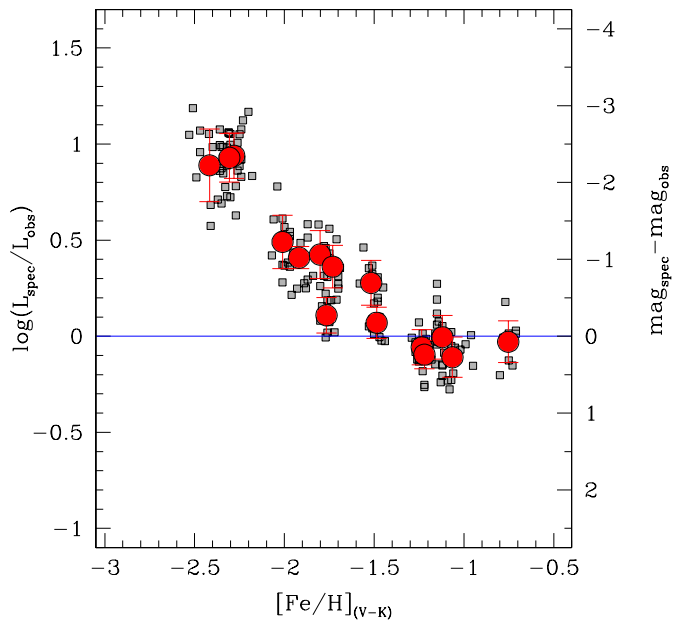}
\caption{Behaviour of the ratio between 
the luminosities derived from spectroscopic and photometric parameters 
as a function of [Fe/H] (the latter derived from 
photometric parameters). The right
vertical axis shows the difference 
in terms of magnitudes.
Same symbols of Fig.~\ref{vialo}.}
\label{lumin}
\end{figure}

As a simple and quantitative test, we compared the luminosities
derived from the observed magnitudes with those 
predicted according to the spectroscopic parameters. 
The stellar luminosity of each target star has been calculated 
both from the observed V-band magnitude as described in Section~\ref{photpars}, and adopting the spectroscopic 
\teff\ and \gr\ . 
Fig.~\ref{lumin} shows the behaviour of the difference between the two luminosities as a function of the metallicity 
(the corresponding magnitude difference is also shown 
in the right vertical axis). 
Spectroscopic parameters predict luminosities 
higher than the observed ones and this difference increases
toward lower metallicities. 
In terms of magnitudes, the most metal-poor GC stars should be $\sim$2.5 magnitudes 
brighter than the observed V-band magnitudes.
This demonstrates that the spectroscopic parameters for metal-poor giant stars are not consistent with 
the evolutionary stage of the stars as inferred from their position in the CMD. 
Hence, the spectroscopic parameters are not reliable locating the stars in 
a wrong position of the \teff\--\gr\ diagram.

\subsection{Technical origin of the parameter discrepancy}

\begin{figure*}
\centering
\includegraphics[width=17cm]{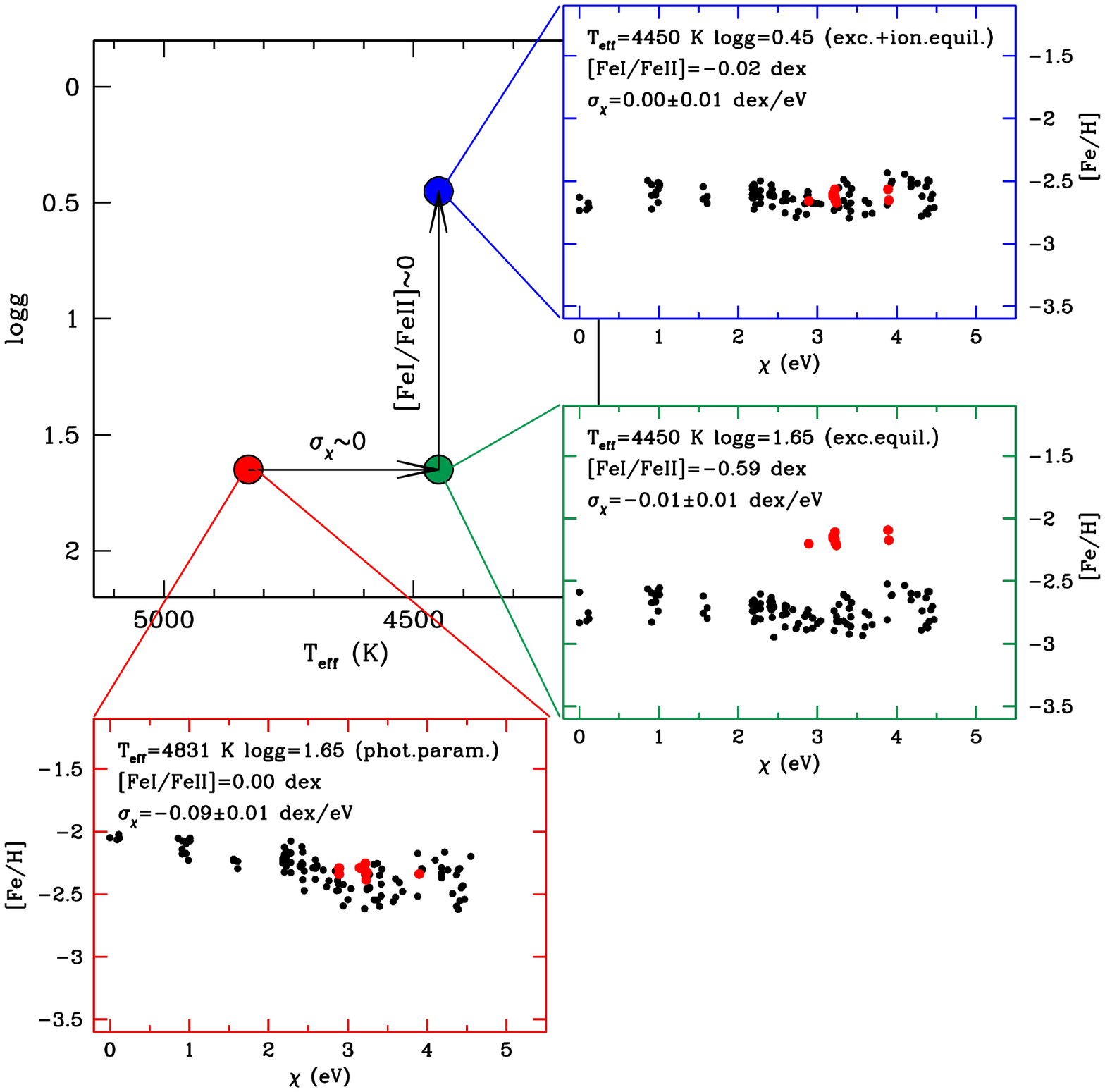}
\caption{Scheme of the location of the star NGC~4590-3584 
in the \teff\--\gr\ plane (main panel) according to 
different chemical analyses: the red circle indicates 
the photometric parameters, the green circle the position 
of the star when the constraint of null \slope\ is 
fulfilled, the blue circle the position of the star according 
to the spectroscopic determination of the parameters. 
For each of these (\teff\ , \gr\ ) pair, the run of 
[Fe/H] as a function of $\chi$ is shown (both neutral and 
single ionised lines, black and red circles respectively).}
\label{momentum}
\end{figure*}

The discrepancy between the spectroscopic and photometric parameters 
is mainly driven by the discrepancy in \teff\ that causes 
those also in \gr\ and \vt\ (and hence in [Fe/H]). 
Fig.~\ref{momentum} explains how a spurious, non-null \slope\ value
leads to a wrong result also in terms of \gr\ and metallicity, locating 
the stars in a wrong position of the \teff\ -\gr\ diagram.

We consider the star NGC~4590-3584 that has photometric parameters 
\teff\ =~4831 K and \gr\ =~1.65 (marked as a red large circle in the \teff\--\gr\ diagram in the main panel of Fig.~\ref{momentum}). 
The photometric \teff\ does not satisfy the excitation equilibrium, providing 
a slope \slope\=--0.09$\pm$0.01 dex/eV, while abundances from neutral and single ionised lines are compatible 
each other ([Fe~I/Fe~II]=~0.0 dex).

In order to null \slope\ , \teff\ needs to be decreased by $\sim$400 K 
(green circle in the main panel). However, a change of \teff\ impacts on both 
Fe~I and Fe~II lines but in opposite directions. In particular a decrease by 100 K 
decreases the abundance from Fe~I lines but increases that by Fe~II lines,
leading to a decrease of [Fe~I/Fe~II] by about 0.18 dex and therefore 
to a decrease of \gr\ by about 0.3 dex. 
In the case of the star shown in Fig.~\ref{momentum}, a decrease of $\sim$400 K 
leads to [Fe~I/Fe~II]=--0.59 dex and we need to decrease \gr\ by 1.2 dex
in order to fulfil the ionisation equilibrium 
(blue circle in the main panel). 

We note that the star NGC4590-3584 lies $\sim$1.8 magnitudes fainter than the RGB Tip but 
it should be located close to the RGB Tip according to its spectroscopic parameters. 
This confirms that the spectroscopic parameters of this star are wrong even if they fulfil 
both ionisation and excitation balance.

\subsection{Physical origin of the parameter discrepancy}

After demonstrating that the spectroscopic parameters are not 
reliable for metal-poor giant stars, we should understand 
the physical origin of this discrepancy. 
The fact that the differences between the 
two sets of parameters increase at lower metallicities suggests that these effects 
are due to inadequacies of the standard model atmospheres/spectral synthesis codes and in 
particular the assumptions of 1D geometry and LTE.

We checked that NLTE effects, under the assumption of 1D geometry, 
are not sufficient to alleviate the parameter discrepancy.
We applied to the Fe~I lines of the three stars shown in Fig.~\ref{exa1} the 
NLTE corrections provided by \citet{bergemann12}\footnote{http://nlte.mpia.de/}. 
We found that for the stars NGC5904-900073 and NGC1904-171 the slopes 
\slope\ do not significantly change, while 
for NGC7099-7414 a non-null \slope\ remains, indicating that 
a significant decrease of \teff\ is requested also with 1D/NLTE abundances.
This finding is compatible with the analysis performed by \citet{amarsi16} 
of the metal-poor giant star HD122563 that shows a similar, negative \slope\ 
both in 1D/LTE and 1D/NLTE (see their Fig.~2). 

On the other hand, 3D effects impact mainly on low $\chi$ lines 
\citep[see e.g.][]{collet07,dobrovolskas13,amarsi16}. 
The star HD122563 discussed by \citet{amarsi16} has parameters and metallicities comparable 
with the most metal-poor stars studied here. 
The 3D/LTE analysis is able to invert the 
observed trend between Fe abundance and $\chi$ providing 
a positive \slope\ . However, the 3D/NLTE analysis provides again a negative 
slope (but less significant than that obtained in 1D/LTE case)
because the NLTE effects counterbalance the 3D effects. 
The results provided by \citet{amarsi16} seem to suggest that a 3D/NLTE analysis 
could partially reduced the discrepancy between the spectroscopic and photometric parameters. 
However this approach still does not provide a flat behaviour between Fe abundance and $\chi$, 
suggesting that our current modelling of 3D/NLTE effects in metal-poor giant stars is still incomplete.

Among the current shortcomings of available 3D model atmospheres we must recall a coarse treatment of opacity (with respect to what possible in 1D model atmospheres) and an incomplete treatment of scattering. Currently 3D model atmospheres either treat scattering as true absorption (e.g. all the older CO5BOLD models of the CIFIST grid \citealt{ludwig09}) or they use an approximate treatment, usually called the {\em Hayek approximation} \citep{hayek10} that consists in treating scattering as true absorption in the optically thick layers and ignore it in the optically thin layers. The effects of the two approximations on the emergent fluxes are discussed in \citet{bonifacio18}. However no investigation has been done of the impact of the different approximations on line formation. We stress that at present no grid of 3D model atmospheres with a full treatment of scattering as done by \citet{hayek10} is available. 
Another possible limitation of the current generation of the two most popular 3D model atmospheres grids, the CIFIST grid \citep{ludwig09} and the STAGGER grid \citep{magic13}, both use the opacity package of the MARCS 1D models \citep{marcs08}, that has been created to compute models 
with effective temperatures below 8000\,K. This implies that there are no opacities for temperatures in excess of 30\,000 K. While such very high temperatures are not encountered
in any layer of 1D models cooler than 8000\,K, in 3D hydro models of cool stars one often finds temperatures that exceed this value, and the codes are obliged to take a bold extrapolation in the opacities.  

Although there is a general consensus that NLTE effects are indeed important in the line formation of metal-poor stars we are also aware that it  the calculations are more complex and must rely on input from atomic physics. Although we believe that the current NLTE computations for Fe are state-of-the art there is still the possibility that there are shortcomings. Among the ones we can think of is that some physical process that may contribute to populate or depopulate atomic levels has either been ignored, or included with an incorrect cross-section (e.g. charge transfer). A common uncertainty is provided by the collisions with neutral hydrogen. The very sophisticated calculation of \citet{amarsi16} did take advantage of quantum mechanical computations for the Fe+H collision rates and included 
charge transfer reactions that lead to Fe$^+$ + H$^-$, the collisions of hydrogen with \ion{Fe}{ii} had to be treated with the unphysical Drawin recipe \citep{drawin68,drawin69}, due to the non-availability of the relevant quantum-mechanical computations. 
Another issue of concern in doing NLTE computations (both in 1D or in 3D) is that the wavelength resolution must be high enough to correctly compute the wings of the strong UV lines that in many atoms effectively control the population. A computation that is too coarse may produce wrong results. 
Of course computations are usually checked against the Sun and Arcturus, however these checks do not guarantee that there will be any shortcoming when computing the line formation in a metal-poor giant. 

A final concern is the possible effects of NLTE on the structure of a 3D model. Both CO5BOLD and STAGGER assume LTE in the model computation, NLTE is taken into account only when computing line formation, using a fixed background model. It is a reasonable assumption, but could be the cause of some shortcoming.

In our view the fact that the most advanced 3D-NLTE computations of \citet{amarsi16} for the metal poor giant HD\,122563 are unable to remove a slope of abundance versus excitation temperature, demonstrates that even using such sophisticated computations, the excitation temperature is unreliable for a metal-poor giant.

\section{A correction scheme for atmospheric parameters}

As  consequence of the above discussion we want to provide ready-to-use empirical recipes that will provide accurate atmospheric parameters of giant stars, that will place them in the correct place in a Hertzsprung-Russel diagram. 

\subsection{RGB stars with [Fe/H]$>$--1.5 dex}
For RGB stars with [Fe/H] $\gtrapprox$--1.5 dex, 
the spectroscopic and photometric approaches are equivalent and the choice of the method is driven 
by the quality of the available photometry and 
spectra. However, the spectroscopic method should be avoided when 
\begin{enumerate}
\item the spectral coverage does not provide a large number of Fe~I lines well distributed 
in $\chi$ and/or line strength, introducing errors in \teff\ and \vt\ ;
\item a few number of Fe~II lines are available, preventing a precise determination of \gr\ .
\end{enumerate}
For these stars, the lines with $\chi<$2 eV can be used because they provide 
abundances coherent with those from high-energy lines, regardless of the used approach 
to derive \teff\ . 

\begin{figure}
\includegraphics[width=\columnwidth]{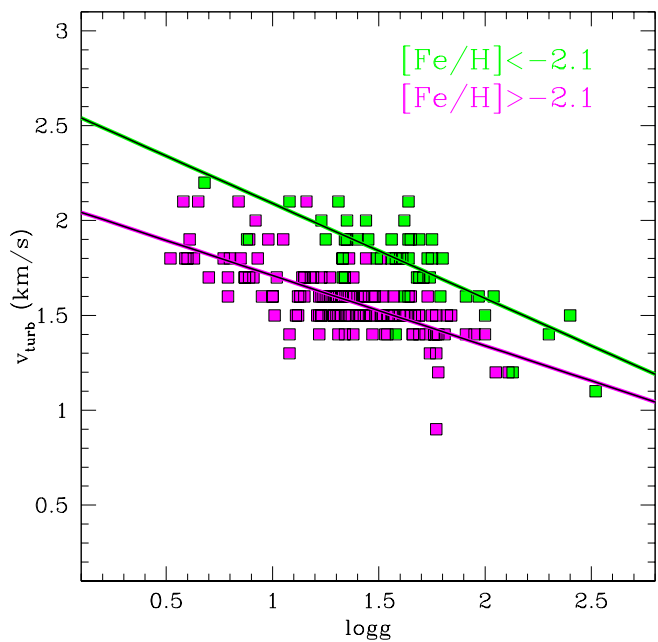}
\caption{Behaviour of \vt\ as a function of \gr\ for the individual stars: 
purple and green squares are the stars in the GCs with [Fe/H]$>$-2.1 dex and [Fe/H]$<$-2.1 dex, 
respectively. Purple and green thick lines are the best linear fits on the two samples of stars.}
\label{myvt}
\end{figure}

Microturbulent velocities must to be derived from the spectra and this parameter is heavily affected by 
the EW distribution of the available Fe~I lines.
As already done by other works \citep[see e.g.][]{monaco05,kirby09}, we provide a linear 
relation between \vt\ and \gr\ in order to 
determine this parameter also in case of 
spectra inadequate to this task. 
As visible in Fig.~\ref{myvt}, where the 
run of \vt\ as a function of \gr\ for all the 
individual stars is shown, there are two 
evident sequences depending on the metallicity. 
For the stars with [Fe/H]$>$--2.1 dex, \vt\ can be 
calculated with the following relation

\begin{align}
{\rm v_{turb}} =   
(-0.37\pm0.03)\cdot{\rm log~g}+(2.08\pm0.04)~  
(\sigma=0.13)
\end{align}

\subsection{RGB stars with [Fe/H]$<$--1.5 dex}
\begin{figure}
\includegraphics[width=\columnwidth]{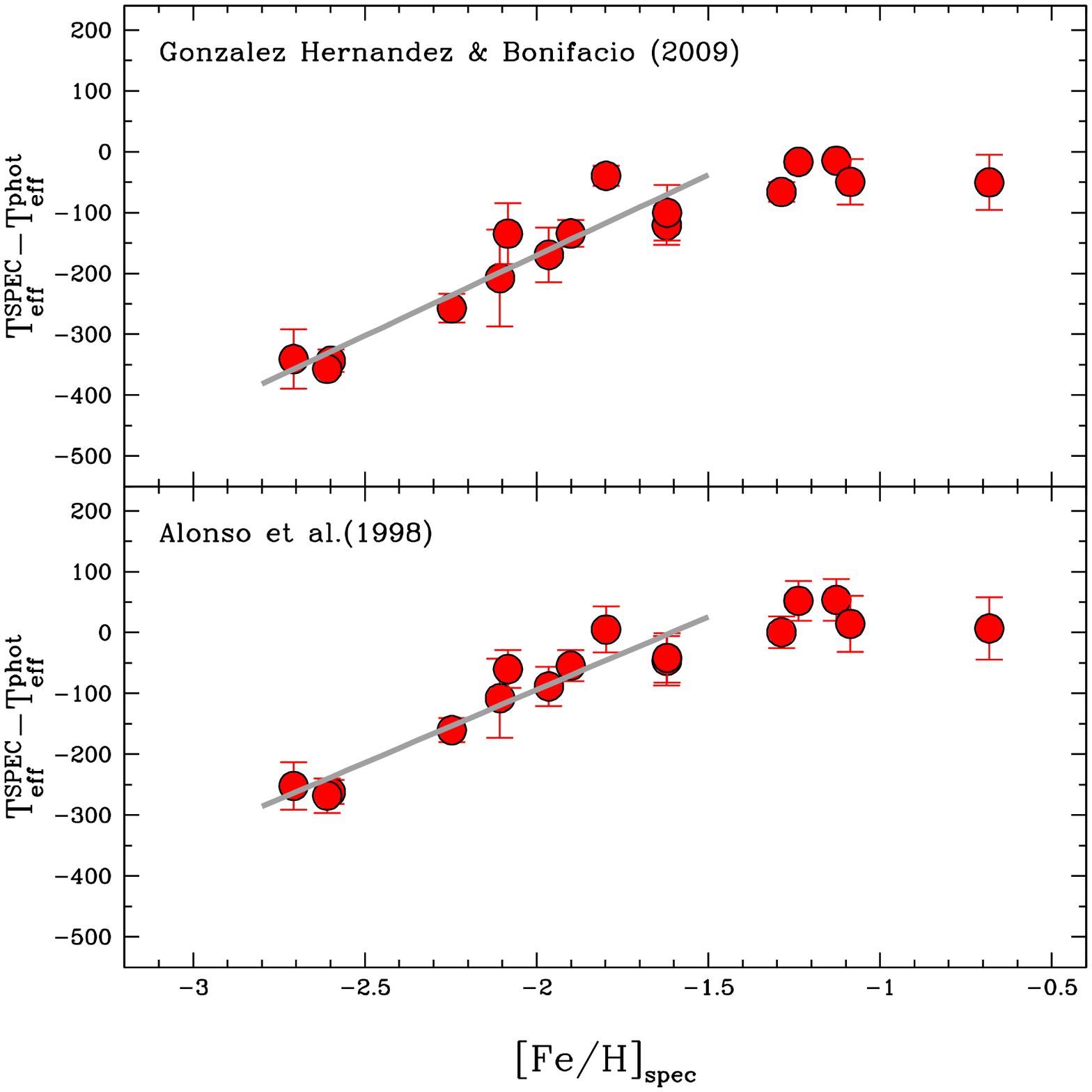}
\caption{
Behaviour of the average difference between 
the spectroscopic and photometric \teff\ obtained 
from individual colours  for target cluster (red circles), as a function of the 
iron abundance derived adopting spectroscopic parameters. 
\teff\ have been obtained with the relations by \citet{ghb09} and \citet{alonso99}, 
upper and lower panel, respectively.
The vertical errorbars are the dispersions of the mean 
of \teff\ for each cluster. Thick grey lines are the 
best linear fits obtained for the clusters with [Fe/H]$<$--1.5 dex.}
\label{linfit}
\end{figure}

For RGB stars with [Fe/H] $\lessapprox$--1.5 dex, 
the photometric approach should be always adopted, even if the available spectra allow a precise determination 
of the parameters. 
Sometime the spectroscopic parameters can be more precise 
(even if less accurate) than the photometric ones because 
of the low quality of the available photometry, the 
uncertainty in the colour excess  or 
the presence of differential reddening. For these cases, 
the spectroscopic parameters can be the only feasible route.
In order to bypass the issues in the spectroscopic parameters discussed above, 
we provide a linear relation between the iron abundance obtained 
with the spectroscopic parameters $\rm [Fe/H]_{spec}$ 
and the average \dteff\ from the broad-band colours, 
both using the relations by \citet{ghb09} and 
\citet{alonso99}, upper and lower panel in Fig.~\ref{linfit} 
respectively.

\begin{align}
{\rm T_{eff}^{GB09}} = T_{\rm eff}^{spec} - 
(264\pm33)\cdot{\rm [Fe/H]_{spec}}-(358\pm70)~  
(\sigma=36 K)
\end{align}

\begin{align}
{\rm T_{eff}^{A99}} = T_{\rm eff}^{spec} - 
(240\pm28)\cdot{\rm [Fe/H]_{spec}}-(385\pm60)~  
(\sigma=31 K)
\end{align}

Note that the relations by \citet{alonso99} 
provide an excellent match with the spectroscopic 
\teff\ for stars with [Fe/H]$>$--1.5 dex , 
while an offset of $\sim$50 K remains when we use 
\citet{ghb09}. On the other hand, \citet{alonso99} 
provide too low \teff\ for the metal-poor clusters 
with respect to theoretical isochrones, while 
\citet{ghb09} provide a good match with the isochrones 
for each metallicity.

Because these relations are defined only for RGB stars 
in the metallicity range --2.5$<$[Fe/H]$<$--1.5 dex, we checked 
whether they work also at lower metallicities. 
We analysed 4 RGB field stars with metallicities between $\sim$--3.5 
and $\sim$--2.5 dex, namely HE~0305-452, CD~38245,      
HD~122563 and HE~2141-3741. For these stars we retrieved archival spectra 
acquired with the spectrograph UVES@VLT, adopting the photometry available 
in the SIMBAD database \citep{wenger00}, colour excess from \citet{sf11} and 
parallaxes from Gaia Data Release 2 \citep{gaia16,gaia18}. 
Photometric and spectroscopic parameters and corresponding [Fe~I/H] 
are listed in Table \ref{field}.
Also for these stars, significant slopes \slope\ are found when the photometric \teff\ 
are adopted, leading to lower spectroscopic \teff\ . 
Additionally, the spectroscopic \gr\ are significantly lower, by about 1 dex, than the photometric ones. 
The precision of the Gaia parallaxes is of about 20\% for HE~0305-452, CD~38245 and HE~2141-3741, and 3\% for HD~122563. However, the precision of the parallax in the first three stars change the photometric gravities by about 0.2 dex and they are not able to 
reconcile photometric and spectroscopic \gr\ .
The spectroscopic \teff\ 
corrected with the relations defined from GCs well match with the photometric \teff\ , 
demonstrating that these relations can be extrapolated at lower metallicities and 
used for very metal-poor RGB field stars (at least down to [Fe/H]$\sim$--3.5 dex), especially when 
no precise photometry and/or colour excess are available.

Concerning the determination of the gravities,
for [Fe/H]$<$--1.5 dex, spectroscopic \gr\ should 
be avoided, because the [Fe~I/Fe~II] ratio is more sensitive 
to \teff\ than to \gr . Hence, if the spectroscopic 
\teff\ is wrong also the spectroscopic \gr\  turns out 
to be wrong, due to the opposite sensitivity of Fe~I and 
Fe~II by \teff\ (see Fig.~\ref{momentum}).
A more robust and safe approach is 
to use the \teff\--\gr\ relation provided by a theoretical 
isochrone (when the age is well known as in the case of a GC) 
or to recalculate gravities adopting the corrected \teff\ .
In this case we give a warning, to derive log g a rough estimate of the mass of the star is needed. If we exclude the cases for which the mass is otherwise known (binary stars, stars with asteroseismic data...), the mass estimate hinges on the age estimate. If we know the star is old (say older than 10 Gyr), as in the case of GCs,  we can safely assume a mass of $\sim$0.7--0.8 M$\odot$. If however the star is younger than 1\,Gyr its mass can be as large as 5\,M$\odot$ leading to a difference of 0.7\,dex in the estimated gravity for the same effective temperature (see Lombardo et al. 2020, in preparation).

Concerning the determination of \vt\ , the stars down to [Fe/H]$\sim$--2.1 dex 
follow the same linear relation provided above, while for the stars in the three most metal-poor GCs (NGC~7078, NGC~4590 and NGC~7099, [Fe/H]$<$--2.1 dex)
we provide the following relation

\begin{align}
{\rm v_{turb}} =   
(-0.50\pm0.06)\cdot{\rm log~g}+(2.59\pm0.10)~  
(\sigma=0.14)
\end{align}

This different behaviour for the most metal-poor stars is due 
to the largest \teff\ discrepancy observed among these stars 
(see Fig.~\ref{vialo}): because the low-$\chi$ lines are the strongest ones, 
\vt\ is increased to partially compensate the negative slope between abundances and $\chi$.

Finally, we note that for these stars the lines with 
low-energy ($<$2 eV) should be used with caution. 
As discussed above, the inclusion of a ten of 
low-$\chi$ Fe~I lines does not significantly 
impact on the average [Fe~I/H] (at least for the optical spectra investigated here where the bulk of the Fe~I lines includes high-$\chi$ lines). 
However, this cut can impact on the abundance 
of other species for which mainly low-$\chi$ could be available. For instance, in the optical range covered by the UVES-FLAMES spectra discussed in this work, almost all the 
Ti lines have $\chi<$2 eV and the adoption 
of a threshold in $\chi$ can dramatically impact on its abundance.

Finally, we recall that Ti provides a significant number of neutral and single-ionised lines, providing an additional diagnostic for the gravities. 
When the spectroscopic parameters are used, and [Fe~I/H] and [Fe~II/H] are consistent within the 
uncertainties by construction,  [Ti~I/H] is lower by about 0.2 dex with respect to 
[Ti~II/H]. This implies that the gravities should be further decreased in order to 
match [Ti~I/H] and [Ti~II/H], worsening the discrepancy with the photometric values. 
We suspect that this behaviour is due to the low $\chi$
of all the available Ti lines. The latter are 
extremely sensitive to the inadequacies of the model atmospheres, in particular 
to NLTE effects, as demonstrated by 
\citet{mashonkina16}, finding that, at [M/H]=--2.0 dex, the NLTE corrections 
for the Ti~I lines are larger than those 
for the Fe~I lines.

\begin{table*}
\caption{Field metal-poor giant stars with the atmospheric parameters and [Fe~I/H] 
derived adopting photometric and spectroscopic parameters.}             
\label{field}      
\centering                          
\begin{tabular}{c c c c c c c c c c c}        
\hline\hline                 
STAR & \teff\ & \gr\ & [Fe~I/H] & \teff\ & \gr\ & [Fe~I/H] & \teff\ & \gr\ & [Fe~I/H]  & PROGRAM \\  
     &        & (A99)&          &        & (GHB09)  &  &   & (SPEC)  &   &       \\    
\hline
HE0305-4520	  &  4801&      1.06&   --3.05&    4896&  1.11&      --2.96&     4300&      0.40&      --3.50&  078.B-0238  \\
HE1116-0634   &  4649&      1.27&   --3.44&    4673&  1.28&      --3.40&     4100&      0.30&      --3.84&  081.B-0900  \\
HD122563      &  4677&      1.37&   --2.71&    4790&  1.43&      --2.60&     4300&      0.40&      --3.03&  065.L-0507  \\
HE2141-3741   &  5100&      1.58&   --3.16&    5217&  1.63&      --3.05&     4650&      0.50&      --3.55&  078.B-0238  \\
\hline                  
\hline                                   
\end{tabular}
\end{table*}

\section{Summary and conclusions}
The analysis of a sample of 16 Galactic GCs observed with UVES-FLAMES@VLT 
using two different approaches to derive the parameters leads to the following results:
\begin{itemize}
\item 
the discrepancy between spectroscopic and photometric parameters for giant stars increases 
decreasing the metallicity. This behaviour is confirmed 
adopting different broad-band colours or colour-\teff\ transformations. 
Such a difference between the two sets of parameters cannot be treated as a simple systematics;

\item
the spectroscopic approach based on excitation and ionisation balances 
provides wrong stellar parameters 
for metal-poor stars, in particular leading to too low \teff\ and \gr\ , inconsistent with the 
values predicted by appropriate theoretical isochrones and with the observed position of the stars in the CMD;

\item
the discrepancy between the two approaches seems to arise from the inadequacies of the adopted physics. 
In particular, low-energy lines are the most prone to 3D effects \citep{bergemann12,dobrovolskas13,amarsi16} 
and the use of 1D model atmospheres is likely 
responsible of the negative values of \slope\ 
that lead to too low \teff\ and \gr\ . 
On the other hand, neither 1D/NLTE  nor
3D/NLTE are  sufficient to flatten the observed \slope\ 
and alleviate the discrepancy between the two parameter sets, at least in the computations currently available;  

\item
we proposed simple relations to correct spectroscopic \teff\ and put them 
onto "photometric" scales by \citet{alonso99} and \citet{ghb09}.
These relations are suitable for the RGB stars with [Fe/H]$<$--1.5 dex 
and they can be used to correct spectroscopic \teff\ both in GCs and 
in field stars when no accurate/precise 
photometry are in hand;

\item
1D (LTE or NLTE) chemical analyses of RGB stars with [Fe/H] $<$ --1.5 dex and based on spectroscopic parameters 
should be considered with great caution 
because the parameters should be underestimated, as well as the derived [Fe/H]. 
We recommend to avoid for these stars 
spectroscopic \teff\ and prefer photometric or 
corrected \teff\ .

\end{itemize}

Finally, we stress that both spectroscopic and photometric \teff\ 
fail to well reproduce that spectral properties of 
giant stars. Spectroscopic \teff\ provide, by construction, 
the same abundance from lines of different $\chi$ but 
clearly fail to reproduce the emerging flux of the stars. 
On the other hand, IRFM photometric \teff\ well reproduce 
the bolometric flux but they provide systematically  erroneous abundances for the low-energy lines. 
In 1D chemical analysis we 
need to decide which aspect we want to privilege, 
a temperature able to reproduce either the emerging stellar flux 
or the depth of individual metallic lines.

Our argumentation about the position of the stars in the \teff\ - \gr\ diagram 
discussed in Section 5 demonstrates that the spectroscopic \teff\ should be rejected, 
while the photometric ones, despite the failure to reproduce 
the excitation balance, are the best choice.

However, the development of more accurate and complete 3D/NLTE tools 
remains the main way to obtain an exhaustive description of 
the stellar spectra and bypass the issues discussed in this work.

\begin{acknowledgements}

We  are  grateful  to  the  anonymous  referee  for  the  useful comments and 
suggestions.

AM is grateful to the Scientific Council of Observatoire de Paris that funded his extended visit at GEPI, 
where part of this work was carried out.

This research has made use of the SIMBAD database,
operated at CDS, Strasbourg, France and 
of data from the European Space Agency (ESA) mission
{\it Gaia} (\url{https://www.cosmos.esa.int/gaia}), processed by the {\it Gaia}
Data Processing and Analysis Consortium (DPAC,
\url{https://www.cosmos.esa.int/web/gaia/dpac/consortium}). Funding for the DPAC
has been provided by national institutions, in particular the institutions
participating in the {\it Gaia} Multilateral Agreement.

\end{acknowledgements}


\end{document}